# Disorder-driven topological phase transition in $Bi_2Se_3$ films

Matthew Brahlek[1,#], Nikesh Koirala[1], Maryam Salehi[2], Jisoo Moon[1], Wenhan Zhang[1], Haoxiang Li[3], Xiaoqing Zhou[3], Myung-Geun Han[4], Liang Wu[5,†], Thomas Emge[6], Hang-Dong Lee[1], Can Xu[1], Seuk Joo Rhee[7], Torgny Gustafsson[1], N. Peter Armitage[5], Yimei Zhu[4], Daniel S. Dessau[3], Weida Wu[1] and Seongshik Oh[1,*]

*Email: ohsean@physics.rutgers.edu

[1]Department of Physics & Astronomy, Rutgers, The State University of New Jersey, Piscataway, New Jersey 08854, USA

[2]Department of Materials Science and Engineering, Rutgers, The State University of New Jersey, Piscataway, New Jersey 08854, USA

[3]Department of Physics, University of Colorado, Boulder, Colorado 80309, USA

[4]Condensed Matter Physics & Materials Science, Brookhaven National Lab, Upton, NY 11973, U.S.A.

[5]The Institute for Quantum Matter, Department of Physics and Astronomy, The Johns Hopkins University, Baltimore, Maryland 21218, USA

[6]Department of Chemistry, Rutgers, The State University of New Jersey, Piscataway, New Jersey 08854, USA

[7]Department of Physics, Hankuk University of Foreign Studies, Yongin-shi, Kyongki-do, 449-791, Korea

[#]Department of Materials Science and Engineering, Pennsylvania State University, University Park, Pennsylvania 16801, USA.

[†]Department of Physics University of California, Berkeley, California 94720, USA

**Abstract: Topological insulators (TI) are a phase of matter that host unusual metallic states on their surfaces. Unlike the states that exist on the surface of conventional materials, these so-called topological surfaces states (TSS) are protected against disorder-related localization effects by time reversal symmetry through strong spin-orbit coupling. By combining transport measurements,**

**angle-resolved photo-emission spectroscopy and scanning tunneling microscopy, we show that there exists a critical level of disorder beyond which the TI $Bi_2Se_3$ loses its ability to protect the metallic TSS and transitions to a fully insulating state. The absence of the metallic surface channels dictates that there is a change in material's topological character, implying that disorder can lead to a topological phase transition even without breaking the time reversal symmetry. This observation challenges the conventional notion of topologically-protected surface states, and will provoke new studies as to the fundamental nature of topological phase of matter in the presence of disorder.**

It has been known since the 1950s that disorder alone can localize electrons and drive metals into an insulating state[1]. Such disorder-driven localization effects are enhanced in 2-dimensions (2D)[2,3]. This is exemplified by the surface states that form on the boundary of non-topological insulators, which are especially susceptible to localization effects by an arbitrarily small level of disorder. In contrast, the novel 2D, Dirac-like topological surface states (TSS) that emerge at the surface of a 3-dimensional (3D) topological insulator (TI) are predicted to be robust to disorder that preserves time-reversal symmetry, in that they should remain metallic[4–6]. So far, experimental efforts to probe the extent of topological protection have been limited to controlling spin-orbit-coupling strength through elemental substitution[7–11] and thickness control in the ultrathin regime of 3D TIs[12–14], where the reduced dimensionality and the overlap of the top and bottom surface wavefunctions are found to compromise the topological protection even without intentional disorder. However, despite theoretical efforts[15–17], the very fundamental question of whether (and how) disorder can drive 3D TIs beyond topological protection remains elusive in real TI materials.

In this work, we address this very issue, by creating the TI $Bi_2Se_3$ films with intentional disorder. As summarized in Table 1, TI phases are generically required to be metallic because of the gapless TSS, whereas metallic nature alone is *not sufficient* to confirm the non-trivial topology because a topologically trivial bulk metal is also metallic, and, therefore, confirmation of a material to be a TI requires other probes such as angle-resolved photo-emission spectroscopy (ARPES). In contrast, however, the material being insulating *is sufficient* to confirm that it is not a TI because a TI is guaranteed to have a metallic surface state[5]. This allows us to show that there exists a critical level of disorder, beyond which the topological $Bi_2Se_3$ transitions into a fully insulating topologically-trivial state. This is achieved by co-depositing Bi and Se onto $Al_2O_3$ substrates at room temperature (~20 °C). At this unusually low growth temperature, rather than being topological and thus metallic, these films are found to be insulating due to the formation of strongly disordered $Bi_2Se_3$ nanocrystals.

To achieve the highest quality single-crystal (referred to herein as low-disorder) $Bi_2Se_3$ films, the optimum growth temperature on $Al_2O_3$ is ~200 – 300 °C[18–21], which gives the adsorbed Bi and Se atoms sufficient surface mobility to diffuse and find an optimal location to crystallize and form single crystals[18]. However, as the deposition temperature is decreased to room temperature and the surface mobility of adatoms decreases, they begin to accumulate in a less coherent manner forming highly disordered nanocrystalline (high-disorder) grains of $Bi_2Se_3$. Further, the disorder can be partially reduced by post-growth annealing of the high-disorder film above 72 °C; these films, which we refer to as an intermediate-disorder state, are found to rearrange structurally to a lower disorder state. Altogether, it is found that three different disorder regimes of $Bi_2Se_3$ are accessible — for more growth details, see Supplemental Information.

Figure 1(a-c) show high-angle annular dark-field scanning transmission electron microscopy (HAADF-STEM) images for films with the different disorder states. The low-disorder sample is clearly single-crystalline with the c-axis orientated out-of-plane. In contrast, the high-disorder state is composed of ~5 nm size grains that are randomly oriented. Finally, the intermediate-disorder sample is composed of nearly c-axis oriented domains with grain size of ~100 nm. X-ray photoemission spectra (XPS) for the Se 3d and Bi 4f in Fig. 1(d-e), respectively, show that despite this large rearrangement, the local chemical environment remains that of $Bi_2Se_3$[22], regardless of the level of disorder. Further, as shown in Fig. 1(f), X-ray diffraction (XRD) measurements indicate a clear change in crystal structure from one with small wide peaks, for the highly disordered films, to the clear c-axis quintuple layer (QL, 1 QL ≈ 0.95 nm) structure for the intermediate (and low) disorder films, which is indicative of the common rhombohedral $Bi_2Se_3$ phase

— a well-known TI. For high disorder films, however, the small peaks are identified as an orthorhombic form of $Bi_2Se_3$[23,24], which is also known to be a TI[25]; the broadness of the peaks and the lack of preferred orientation implies that this phase is composed of disordered nanocrystals, which is consistent with the HAADF-STEM images above. Altogether, HAADF-STEM, XPS and XRD show that all the films are composed of $Bi_2Se_3$ phases at the nanoscale that are known TIs, but, as we will see below, the level of disorder manifests dramatically in the electronic properties.

Figure 2(a-c) shows resistance versus temperature measurements for low, intermediate and high disorder films, respectively. The resistance for both low- and intermediate-disorder films decreases with decreasing temperature, which is indicative of metals. For the high-disorder state, however, the room temperature resistance is three orders of magnitude larger, and the resistance quickly increases with decreasing temperature to a point where it becomes immeasurably large below ~150 K: in other words, it is an insulator. As shown in Tab. 1, the observation of the insulating behavior directly indicates that the high-disorder phase is non-topological. If the sample were topological, the interface with vacuum (or air) should have metallic surface states. Therefore, the very fact that the high-disorder $Bi_2Se_3$ is not metallic implies that it is non-topological. However, the insulating nature and the lack of long-range-order (due to small randomly oriented grains) precludes (see Supporting Information) other probes such as ARPES and scanning tunneling microscopy (STM) from proving the trivial topology of this state, which makes transport measurements the only (but still sufficient) probe showing that the highly disordered state of $Bi_2Se_3$ is non-topological.

In contrast to the insulating phase, the observation of metallic transport shown in Fig. 2(a-b) is not sufficient to confirm the topological nature of these states. This is due to the fact that carriers can be transported through the bulk, through the TSS or through both. Therefore, other probes are required to confirm whether a metallic system is topological or not. Fig. 2(d-e) show ARPES measurements for low- and intermediate-disorder phases, respectively. For the low-disorder film, the bulk band and the Dirac-like TSS are clearly visible, which confirms that the sample is topological. For the intermediate-disorder films, however, the spectrum is diffuse and much broader with noticeable suppression of the intensity at the Dirac point, and, thus, inconclusive regarding the topology of this sample. The beam size for APRES experiments performed here is large (> 100 μm) compared to the nanoscale microstructure of the films in this study. Therefore, the smearing of the spectrum in Fig. 3(e) may be due to the superposition of grains with disparate electronic structures.

To understand how the local electronic properties affect the macroscale properties measured by transport and ARPES, Fig. 2(g-l) shows local topographical maps and tunneling spectroscopy for low- (STM), intermediate- (STM), and high-disorder (atomic force microscopy) films, respectively. In Fig 2(g-i) there is a clear difference in surface morphology between different levels of disorder, and are consistent with the STEM and XRD data shown in Fig. 1. The low-disorder sample shows clear large quintuple layer (QL, the minimum unit of $Bi_2Se_3$) terraces typical of epitaxial $Bi_2Se_3$. The intermediate-disorder film shows both ordered QL step terraces and disordered areas. In contrast, there are not any long-range crystalline features visible on the high-disorder sample in Fig. 2(i), which is consistent with the STEM image in Fig. 1(c). Further, scanning tunneling spectroscopy (STS) measurements enable probing the electronic structure in a spatially resolved way. Fig. 2(j-k) show STS measurements for low- and intermediate-disorder samples, respectively. For the low-disorder sample the differential conductance, $dI/dV$, decreases with negative bias, until it reaches a minimum around -0.4 eV, at which point it again starts to increase. This is typical of the Dirac surface band as observed many times before[26]. For the intermediate-disorder sample, the STS spectrum from the terraced area is nominally identical to the low-disorder STS data. This indicates that there exists TSS and, at least locally, the material is topological. However, the more disordered regions in Fig. 2(h) was so insulating that STS spectrum was not measurable. This suggests that either the more disordered region does not contain TSS (non-topological) or the TSS exists below vacuum-$Bi_2Se_3$ interface in this region. This observation shows that the intermediate-disorder sample may be phase-segregated into topologically trivial and non-trivial regions. It is worth noting that despite the similarities between the transport properties between the low-disorder and intermediate-disorder samples, their ARPES and STM (STS) data are vastly different. This is in fact easy to understand considering that transport properties are dominated by the most conducting paths of the sample as far as the conducting paths are contiguous throughout the sample, whereas ARPES requires long-range crystalline order and STM (STS) probes local properties.

Fig. 2(m) shows a phase diagram summarizing the results for the evolution of a TI phase with increasing disorder. Under the $Z_2$ TI paradigm, all non-magnetic insulators can be grouped into either topological insulators, which are guaranteed to have metallic surface states, or trivial insulators, whose surface states are not guaranteed to be metallic. According to this classification, the disordered insulator phase studied here should belong to the trivial insulator. However, it should be noted that disorder itself is a non-specific property, covering a broad range of material states that interrupts the real-space periodicity of the underlying lattice in combination of point, linear, or planar defects. The question is then: what specific aspect of disorder drives a TI to a trivial insulating state? In our present case the disorder is in the form of decreasing grain size with random orientation, which ultimately must be the driver that changes the material's global topology.

It is well known that near a phase transition small local fluctuations in disorder can dramatically shift the global ground state. This has long been studied from percolation theory near a metal-insulator phase transition[27], electron-hole phase segregation near the Dirac point in graphene[28], to fluctuating anti-ferromagnetism that may stabilize high-temperature superconductivity[29]. The $Z_2$ classification scheme should be no exception, and must be sensitive to local disorder[30]. Our present study shows that topological-percolation most likely drives the observed phase transition from a TI to a non-TI, and, therefore, the transition cannot be considered abrupt in the disorder phase space. As STM shows, in the intermediate-disorder sample the material is phase-segregating into regions with and without gapless surface states, while globally the material remains sufficiently well connected as to remain metallic. As disorder is further increased with grain sizes reducing from ~100 nm to ~5 nm scale, non-topological regions grow and topological regions shrink, and eventually the materials become globally insulating when the metallic topological islands are sufficiently surrounded by small trivially insulating grains. Such a combination of topological-disorder and percolation likely plays an important role even in other topological transition materials such as $(Bi_{1-x}In_x)_2Se_3$[8,31,32]. Still, with percolation aside, it remains an open question what drives the local disordered regions to a net-insulating, non-topological state as their grain size decreases. Instead of attempting to provide a definite answer to this question, below we present a simple physical picture as a guide to future studies.

The transition away from the TI phase observed here likely comes down to a combination of the size of the grains, their random orientation with respect to each other, and the details of the interfaces. Considering the grain size, it is well known that making a TI film thinner than a critical thickness (~4-5 nm)[12] opens a gap at the Dirac point, and hence, drives the TI into a trivial state. This results from the TSSs on the top surface hybridizing with the TSS on the bottom surface since they have opposite spin direction. In terms of finite thickness, the small grain size of the high-disorder film is similar to the 4-5 nm critical thickness. This should, however, not compromise the topological properties of the material because finite thickness effects should only occur at the interface of a TI and non-TI when the TI is below this critical thickness.

The grain boundaries are composed of random crystallographic surfaces and how this affects the long range properties of the materials is not clear. As the grains are made smaller, the reciprocal lattice vector loses meaning and energy bands flatten, with the *k*-dependent band structure being replaced by a conglomeration of local molecular orbitals. Defining a $Z_2$ topological invariant only relies on the presence of an energy gap, and does not require *k*-space being well defined[33]. Reducing the size of crystalline grains, however, is well known to only increase the energy gap due to quantum confinement effect, which should not cause a global band inversion requisite for a TI-to-non-TI phase transition. This then prompts the question of how the systematic increase in crystalline disorder kills the global topological invariant as observed here. To better answer this question, future explorations into the materials properties that control grain size will be required; this will provide a means to explore the nature of the transition from a topological insulator to a disordered trivial insulator in a continuous way. Altogether, the current study raises the challenging theoretical and experimental questions as to how to define and probe topological nature of disordered materials.

**Acknowledgements**

We would like to thank Emil Prodan for helpful discussion. This work is supported by NSF (DMREF-1233349, DMR-1126468, DMR-1308142, DMR-1506618 and EFMA-1542798), Gordon and Betty Moore Foundation's EPiQS (GBMF4418) for SO, DOE (DE-FG0203ER46066) for DSD and Gordon and Betty Moore Foundation (GBMF2628) for NPA. The work at Brookhaven National Lab is supported by U.S. Department of Energy, Office of Basic Energy Science, Division of Materials Science and Engineering under Contract number DE-AC02-98CH10886. TEM sample preparation was carried out at the Center for Functional Nanomaterials, Brookhaven National Laboratory. The Advanced Light Source is supported by the Director, Office of Science, Office of Basic Energy Sciences, of the U.S. Department of Energy under Contract No. DE-AC02-05CH11231.

**Captions**

**Table 1.** The relationship between transport properties being metallic or not and the material being topological or not. If a material is metallic, then it could be topological or non-topological. However, if a material is insulating then it cannot be topological because a topological material is guaranteed to have a metallic surface state.

**Figure 1.** Structural and chemical comparison of disordered $Bi_2Se_3$ thin films with various levels of disorder. (a-c) High-angle annular dark-field scanning transmission electron microscopy images for films with low-, intermediate- and high-disorder, respectively. The low-disorder one (a) shows fully single crystalline ordering and the intermediate one shows some grain boundaries with ~100 nm grain sizes but with still mostly c-orientation. On the other hand, the high-disorder film (c) is composed of ~5 nm crystalline grains (outlined by dashed lines) without any preferential orientations. (d-e) X-ray photo-emission spectra show that the Se-3d (d) and Bi-4f (e) levels are independent of the level of disorder. (f) X-ray diffraction data show that the high-disorder films is composed of orthorhombic $Bi_2Se_3$ disordered grains, whereas intermediate- and high-disorder are formed of rhombohedral $Bi_2Se_3$ phase with dominant c-orientation: note the crystal orientation indices.

**Figure 2.** Electrical transport, and spectroscopy measurements as a function of disorder strength (Low-disorder: left column, intermediate-disorder: center column, and high-disorder: right column). (a-c) Resistance versus temperature. (d-f) Angle-resolved photoemission spectroscopy (ARPES). ARPES for the high disorder sample is not possible due to lack of long range crystalline order, combined with its insulating behavior. (g-i) Surface topography. The low-disorder and intermediate-disorder were performed with scanning tunneling microscopy ($V_B$ = -2 V, $I_T$ = 5 pA), whereas, due to the highly insulating properties, the high-disorder case was performed with atomic force microscopy. (j-k) Scanning tunneling spectroscopy, taken from X-marked regions of (g) and (h). For the intermediate-order sample, STS was not possible in the more disordered regions away from the X-mark. STS for the high-disorder sample was not possible due to its insulating behavior. (m) A phase diagram summarizing the electronic phase as a function of disorder.

**Tab. 1 (single column)**

| | Insulating | Metallic |
|---|---|---|
| **Topological** | × | ✓ |
| **Non-topological** | ✓ | ✓ |

**Figure 1. (double column)**

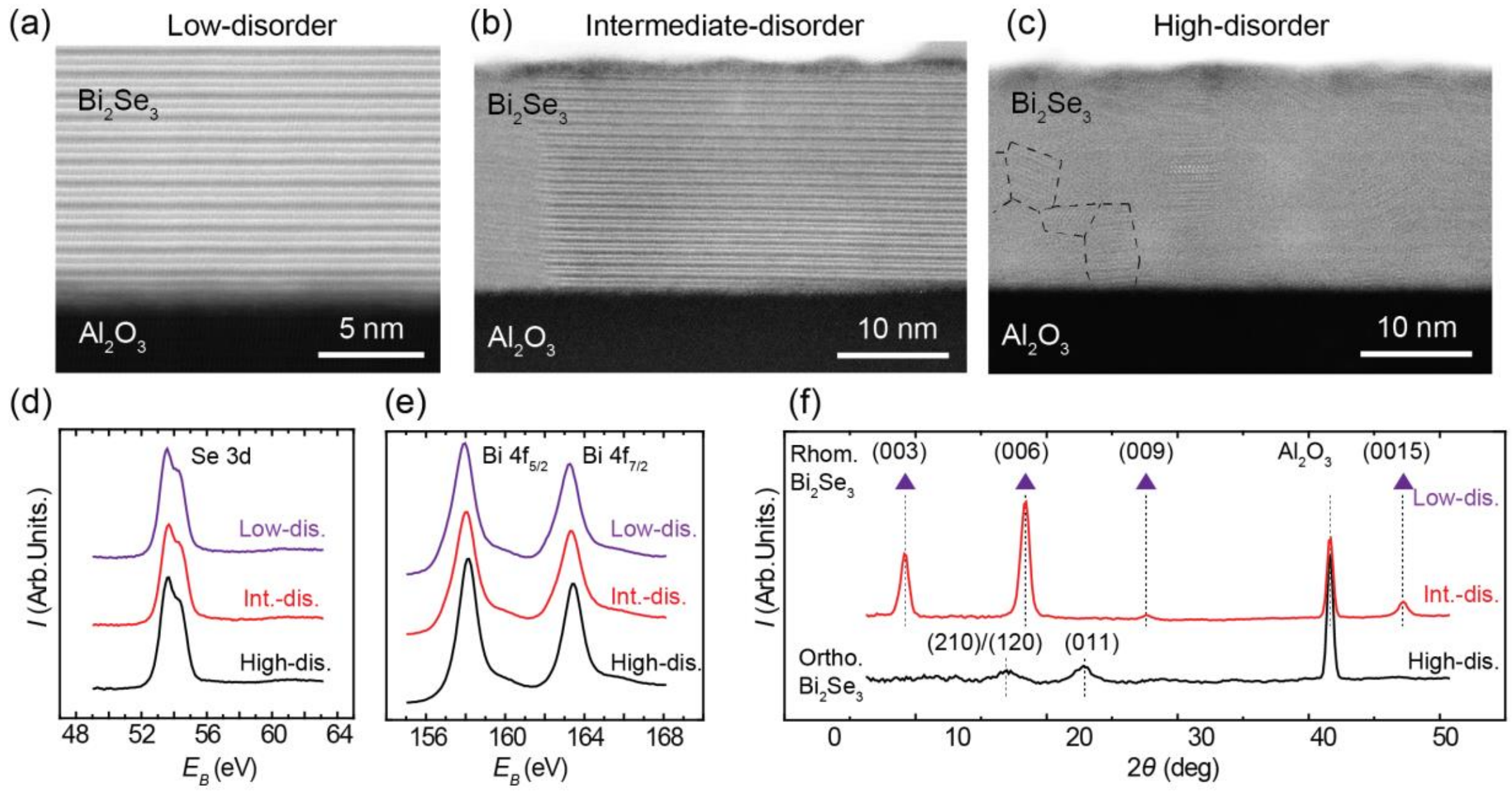

**Figure 2. (single column)**

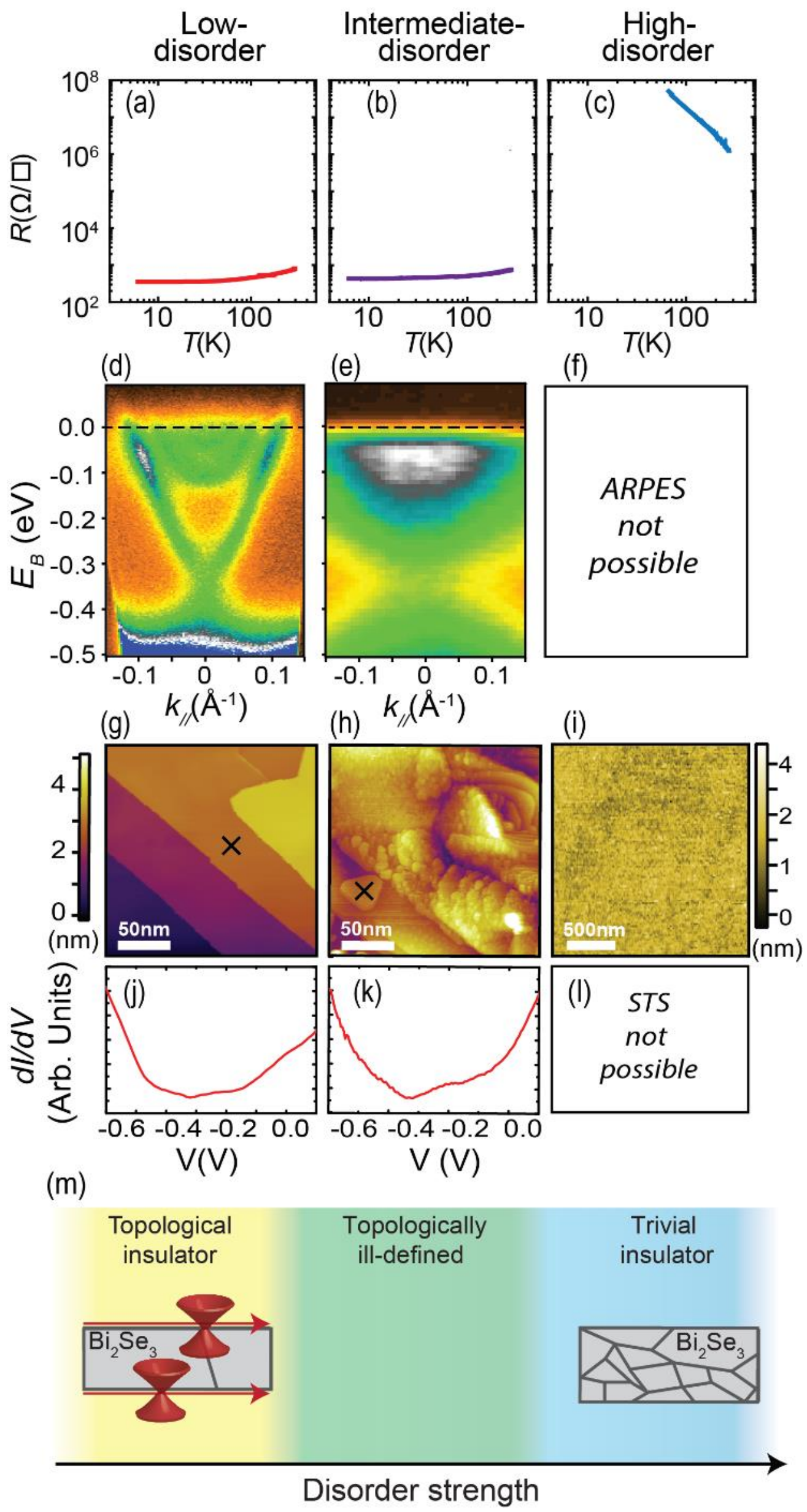

Supplemental Information for

- S1 – Experimental Methods
- S2 – Se-vacancy Suppression

**S1-Experimental Methods**

The films used in the work were grown by molecular beam epitaxy (MBE, SVT-Associates, Inc), with standard thermal cells for Bi and Se. The substrates used were 10 mm × 10 mm × 0.5 mm $Al_2O_3$ (0001), which were cleaned *ex situ* for 5 minutes with ozone prior to being vacuumed down in the MBE chamber. Once inside, the substrates were further cleaned by heating to 750 °C (as measured by a thermocouple mounted behind the substrate) in molecular oxygen for ~10 minutes. To grow the low-disorder films, deposition took place using the two-step method[1]: the substrates were cooled to 135 °C where 3 quintuple layers (QL, 1 QL ≈ 0.95 nm) $Bi_2Se_3$ were deposited, which was followed by slow heating to 300 °C where another 17 QL (a total of 20 QL) was deposited. During growth the Se to Bi flux ratio was set to be ~10:1 to minimize Se-vacancies; after growth, the sample was cooled in constant Se flux also to minimize Se-vacancies.

For the high- and intermediate-disorder samples the substrates were cooled to room temperature (~20 °C) where the deposition took place to a thickness of 20 QL. To obtain a consistent deposition temperature the substrates were left at room temperature overnight. Prior to each deposition the Bi and Se cells were idled at least 2 hours before being calibrated to a ratio of Se/Bi ≈ 1.7 by an *in situ* quartz crystal microbalance (QCM), which was confirmed by *ex situ* medium energy ion scattering (MEIS). After deposition, the films were either removed (high-disorder) with no subsequent annealing or annealed (intermediate-disorder) at the desired temperature. To anneal the films in a controllable and reproducible manner, the substrate heater was controlled with a standard PID (proportional-integral-differential) control loop. The ramp rate was chosen to be ~1 °C/min, which was used to avoid overshooting the desired temperature and to maintain reproducibility; all films labeled as intermediate disorder were annealed above 72 °C. Once the desired temperature was reached, the heating was immediately stopped and the films cooled down while being removed from the sample stage.

Transport measurements were performed in the standard square van der Pauw geometry. Indium leads were pressed at the corners, which formed Ohmic contacts for both metallic and insulating samples. Once electrical contact was made the samples were cooled to ~6 K, where, if the samples were metallic, magneto-resistance measurements were performed, with a ~0.6 T electro-magnet with the field directed perpendicular to the sample surface. The magneto-resistance data were then symmetrized with respect to the magnetic field, and the carrier density was extracted from the slope of the transverse component $\Delta R/\Delta B$, as $n_{2D} = (e(\Delta R/\Delta B))^{-1}$.

Samples used in scanning tunneling microscopy (STM) and angle-resolved photo-emission spectroscopy measurements (ARPES) were grown at 300 °C (low-disorder) or at 20 °C and annealed above 72 °C (intermediate-disorder) and capped in situ with ~100 nm Se at room temperature. Following the decapping procedure outlined in Ref. [2], the Se layer was removed by brief ion milling followed by thermal desorption at > 200 °C. Both STM and ARPES were not possible on the high-disorder samples for the following reasons: (1) The samples could not be decapped without transitioning to the intermediate-disorder state, due to the thermal desorption step at > 200 °C. (2) Even if the surface protection were not a technical challenge, these samples were too insulating to perform either ARPES or STM measurement. (3) ARPES requires reasonably well-defined crystal momentum for it to work as a probe, and, therefore, the complete lack of crystal-momentum in the high-disorder sample disallows any meaningful ARPES data to be produced after all. Considering these factors, the only reliable probe to detect the electronic state in this highly insulating regime is the transport measurement.

ARPES was performed both with laser and synchrotron based light sources. The laser ARPES was done using 7 eV light, measured with true pulse counting on a SPECS analyzer. The synchrotron data was taken at BL10 at the Advanced Light Source, Berkeley. All ARPES was performed in UHV ($< 1 \times 10^{-10}$ Torr) vacuum chambers following in-situ decapping of the Se overlayers.

STM was performed with an Omicron UHV-LT-STM system at a base pressure less than $2 \times 10^{-11}$ Torr. The differential conductance measurements were carried out with a standard lock-in technique with an amplifier gain $R_{Gain} = 3$ GΩ, a modulation frequency $f = 455$ Hz, and an amplitude $V_{Mod} = 10$ mV.

The sample preparation for TEM was carried out with a focused-ion beam (FIB) setup using 5 keV Ga+ ions. A JEOL ARM 200CF equipped with a cold field-emission gun and double-spherical aberration correctors operated at 200 kV was used for high-angle annular dark-field (HAADF) scanning transmission electron microscopy (STEM) with the collection angles ranging from 68 to 280 mrad.

**S2-Se-vacancy Suppression**

$Bi_2Se_3$ is well known to easily form Se vacancies, which contribute charge carriers. During the growth at 300 °C the adatom surface mobility is very high, which allows excess Se to desorb leaving only a minimal number of Se vacancies which is limited by thermodynamics. However, at 20 °C the adatom surface mobility is low, which makes it difficult for Se to diffuse around and quench local Se vacancies. Therefore, in order to fully reduce the Se vacancies, we had to use slightly excess Se at Se/Bi of ~1.7. This effect can be seen in Figure S1, which plots the carrier concentration and mobility for $Bi_2Se_3$ films as-grown at 20 °C as a function of varying Se/Bi ratios. The excessive Se vacancies lead to much higher sheet carrier density ($10^{14}$~$10^{15}$ /cm$^2$) in these films than that (~$10^{13}$ /cm$^2$) of the standard low-disorder TI films, even up to Se/Bi of ~1.6. As the excessive Se vacancies are suppressed with extra Se, the carrier densities decrease and become immeasurably small for Se/Bi of ~1.7 or higher. It is worth noting that while the carrier density decreases monotonically with increasing Se content, the mobility is less affected, suggesting that the vacancy removal affects mostly the carrier density but not the level of disorder.

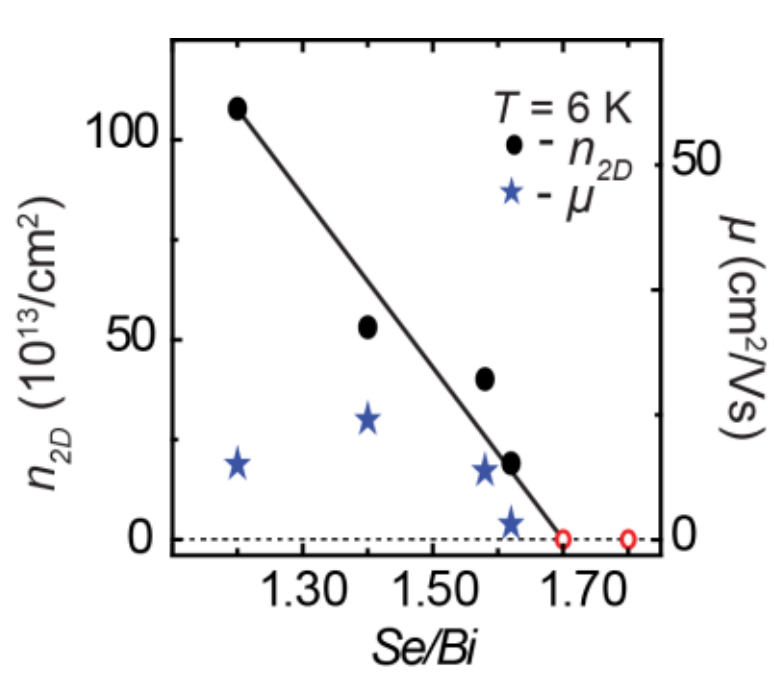

**Figure S1.** Sheet carrier density (left axis) and mobility (right axis) measured at 6 K. The carrier density decreases with increasing Se/Bi ratio, and vanishes between 1.6 – 1.7, as shown by the guiding line. The red open circles represent the carrier densities for Se/Bi of 1.7 and 1.8, which are immeasurably small, thus taken as zero.